\documentstyle[aps]{revtex}
\draft
\begin{document}
\title{Generator coordinate method calculations of one-nucleon removal
reactions on $^{40}$Ca}
\author{M.V. Ivanov, M.K. Gaidarov, and A.N. Antonov}
\address{Institute of Nuclear Research and Nuclear Energy,
Bulgarian Academy of Sciences, Sofia 1784, Bulgaria}
\author{C. Giusti}
\address{Dipartimento di Fisica Nucleare e Teorica,
Universit\`a di Pavia,\\
Istituto Nazionale di Fisica Nucleare, Sezione di Pavia,
Pavia, Italy}
\maketitle

\begin{abstract}
An approach to the Generator Coordinate Method (GCM) using
Skyrme-type effective forces and Woods-Saxon construction
potential is applied to calculate the single-particle proton and
neutron overlap functions in $^{40}$Ca. The
relationship between the bound-state overlap functions and the
one-body density matrix has been used.  These overlap functions
are applied to calculate the cross sections of one-nucleon
removal reactions such as ($e,e'p$), ($\gamma,p$) and ($p,d$) on
$^{40}$Ca on the same theoretical footing. A consistent description
of data for the different reactions is achieved.
The shapes of the experimental cross sections for transitions to the
$3/2^{+}$ ground state and the first $1/2^{+}$ excited state of the
residual nuclei  are well reproduced by the overlap functions obtained
within the GCM. An additional spectroscopic factor accounting for
correlations not included in the overlap function must be applied to
the calculated results to reproduce the size of the experimental
cross sections.

\end {abstract}
\vspace{1cm}

\section{Introduction}
Experiments on nuclear reactions accompanied by one-nucleon
removal from $^{40}$Ca (e.g. \cite{Roos75,Kra90,Lap93,Abe92}) have
accumulated much spectroscopic information on its nucleon-hole
spectral density function and, generally, on the single-particle
aspects of nuclear structure. From the theoretical point of view
two topics in the analyses of these processes are of significant
interest and have been mainly studied: the reaction mechanism and
the ground-state correlation effects. The latter can be
successfully considered by using the unique relationship between
the overlap functions (OF) related to bound states of the
$(A-1)$-particle system and the one-body density matrix (ODM) of
the $A$-particle system in its ground state \cite{Vn93}. This
makes it possible to investigate the effects of the various types
of nucleon-nucleon correlations included in the ODM on the
bound-state proton and neutron overlap functions.

In our recent works \cite{Gai99,Gai2000} a consistent study of
overlap functions and one-nucleon removal reactions on $^{16}$O
using different correlation methods has been carried
out and the comparison with the experimental data has been
performed. In our previous calculations we used methods which
account mainly for short-range and tensor nucleon-nucleon
correlations. It is desirable, however, to take into
consideration also correlations originating from the collective
motion of the nucleons. This was partially done for the $^{16}$O
nucleus in \cite{Gai2000}. In this respect, the various
applications of the Generator Coordinate Method to nuclear
problems \cite{Chr86,Ant,Iva2000} have shown its efficiency as a
potential source of information on nucleon-nucleon correlations in
nuclei. The results on the one- and two-body density and momentum
distributions, occupation probabilities and natural orbitals
obtained within the GCM using various construction potentials
\cite{Iva2000} have shown that the nucleon-nucleon correlations
accounted for in this method are different from the short-range
ones and are rather related to the collective motion of the
nucleons. It was pointed out that these correlations are also
important in calculations of other single-particle properties,
such as one-body overlap functions, which are necessary in the
calculations of cross-sections of one-nucleon removal reactions.

It is known that, in general, when going from light to heavier
systems the collectivity becomes stronger. Because of its
undeformed closed-shell structure, the medium-heavy $^{40}$Ca
nucleus is one of the few nuclei for which microscopic
calculations can be performed. For instance, only recently the
Fermi hypernetted chain theory has been extended to describe the
ground-state properties of $^{40}$Ca \cite{Sa96}. Therefore, it
is very attractive to probe its "doubly-magic" character also in
calculations of different types of one-nucleon removal reactions.
As a first step differential cross sections for the
$^{40}$Ca($p,d$) pick-up reaction have been calculated in
\cite{Di97} with overlap functions obtained from the ODM in the
Jastrow correlation method (JCM) \cite{Sto96}. Although only
short-range correlations have been included in the OF, it was
shown that the angular distributions obtained are in a qualitative
agreement with the empirical ($p,d$) data for the transition to
the ground state of the residual nucleus. These results have
clearly pointed out the necessity of inclusion of another kind of
NN correlations, namely the long-range correlations,
corresponding to collective degrees of freedom. Better agreement
with the experimental $^{16}$O($e,e'p$) and $^{16}$O($\gamma,p$)
data was achieved in \cite{Gai2000}, where it was concluded that
the long-range correlations affect the spectroscopic factors
causing an additional depletion of the quasihole states.

The main aim of the present work is to study the effects of the
nucleon-nucleon correlations included in the Generator Coordinate
Method on the behaviour of the bound-state proton and neutron
overlap functions in $^{40}$Ca and of the related one-nucleon
removal reaction cross sections. We first calculate both proton
and neutron single-particle overlap functions and spectroscopic
factors on the basis of the corresponding proton (with included
Coulomb interaction) and neutron one-body density matrices of the
$^{40}$Ca nucleus obtained with GCM using the relationship
between the ODM's and the overlap functions. Second, these
bound-state OF are used to calculate the cross sections of the
$^{40}$Ca($e,e'p$), $^{40}$Ca($\gamma,p$) and $^{40}$Ca($p,d$)
reactions. Thus it becomes possible to investigate the role of
the correlations related to collective nucleon motions and
accounted for in the GCM on the overlap functions and one-nucleon
removal cross sections in $^{40}$Ca also in comparison with data.

In Section II we give the basic theoretical relationships
necessary to obtain the one-body density matrix in GCM and the
bound-state overlap functions by means of the asymptotic
behaviour of ODM. The results of the calculations of overlap
functions and cross sections of ($e,e'p$), ($\gamma,p$) and
($p,d$) reactions on $^{40}$Ca are presented and discussed in
Section III. Section IV contains the concluding remarks.

 \section{The theoretical scheme}
\subsection{The GCM  ground-state one-body density matrix}
We start from a standard GCM-type $A$-particle  wave function
$\Psi$ when one real generator coordinate $x$ is considered
\cite{Gri57}, i.e.
\begin{equation}
{\Psi}(\{{\bf{r}}_i\})=
\int_{0}^{\infty}{f(x){\Phi}(\{{\bf{r}}_i\},x)dx}\,,\;\;i=1,...,A.
\label{eq:psi}
\end{equation}
In Eq. (\ref{eq:psi}) ${\Phi}(\{{\bf{r}}_i\},x)$ is the generating
function, $f(x)$ is the generator or weight function and $A$ is
the mass number of the nucleus.

The application of the Ritz variational principle ${\delta}E=0$
leads to the Hill-Wheeler integral equation \cite{Gri57,Hil53} for
the weight function and the energy of the system:
\begin{equation}
\int_{0}^{\infty}{[{\cal H}(x,x')-EI(x,x')]f(x')dx'}=0\,,
\label{eq:HW}
\end{equation}
where
\begin{equation}
{\cal H} (x,x')=\langle {\Phi}(\{{\bf{r}}_i\},x)|\hat{H}|{\Phi}
(\{{\bf{r}}_i\},x')\rangle\,
\label{eq:ek}
\end{equation}
and
\begin{equation}
I(x,x')=\langle
{\Phi}(\{{\bf{r}}_i\},x)|{\Phi}(\{{\bf{r}}_i\},x')\rangle \,
\label{eq:ok}
\end{equation}
are the energy and overlap kernels, respectively, and $\hat{H}$
is the Hamiltonian of the system. Solving the Hill-Wheeler
equation (\ref{eq:HW}) one can obtain the solutions
${f}_0,{f}_1,...$ for the weight functions which correspond to
the eigenvalues of the energy ${E}_0,{E}_1...$.

The corresponding ground-state one-body density matrix is given
by (see, e.g. \cite{Ant}):
\begin{equation}
\rho({\bf{r}},{{\bf{r'}}})=\int\!\!\int
{f}_{0}(x){f}_{0}(x')I(x,x'){\rho}(x,x',{\bf{r}},{\bf{r'}})dxdx'
\label{eq:obdm}
\end{equation}
with
\begin{equation}
{\rho}(x,x',{\bf{r}},{{\bf{r'}}})=4\sum_{{\lambda},{\mu}=1}^{A/4}
{{{(N^{-1}(x,x'))}_{{\mu}{\lambda}}}{{\varphi}_{\lambda}^{*}}({\bf{r}},x)
{{\varphi}_{\mu}}({{\bf{r'}}},x')}\,,
\label{eq:nmatr}
\end{equation}
where $\varphi_{\lambda}({\bf r},x)$ are the single-particle wave
functions corresponding to a given construction potential by means
of which the generating single Slater determinant wave function
${\Phi}(\{{\bf{r}}_i\},x)$ is built. The matrix
$N_{\lambda\mu}^{-1}(x,x^{\prime })$ in Eq. (\ref{eq:nmatr}) is
the inverse matrix of
\begin{equation}
{N}_{{\lambda}{\mu}}(x,x')=
\int_{}^{}{{\varphi}_{\lambda}^{*}({\bf{r}},x){\varphi}_{\mu}({\bf{r}},x')}
d{\bf{r}}\,.
\label{eq:lamu}
\end{equation}

As known, the results of the GCM calculations depend on the type
of the construction potential used to define the single-particle
wave functions $\varphi_{\lambda}({\bf r},x)$. In the present
work we choose the Woods-Saxon (WS) potential as a construction
potential with the diffuseness parameter as a generator
coordinate. The GCM scheme, which has been already adopted in
\cite{Gai2000}, gives a very good agreement with the data for the
$^{16}$O($e,e'p$) and $^{16}$O($\gamma,p$) reaction cross
sections. In our present calculations the Skyrme-type effective force
$SkM^{*}$ \cite{Bar82} is used, with parameters which give
realistic binding energy of $^{40}$Ca obtained from the
Hill-Wheeler equation. The agreement of the calculated
basic nuclear characteristics with their empirical values
obtained in \cite{Iva2000}, proved once more the reliability of
these effective forces, which are used also in the present study.

\subsection{The overlap functions for the proton and neutron bound states}
For a correct calculation of the cross section of nuclear
reactions with one-neutron or one-proton removal from the target
nucleus, the corresponding overlap functions for the neutron and
proton bound states must be used in the reaction amplitudes. The
construction of the ground-state ODM of $^{40}$Ca from the
Generator Coordinate Method calculations makes it possible to
apply the procedure for extracting single-particle overlap
functions \cite{Vn93}. Here we would like to remind that the
single-particle overlap functions are defined by the overlap
integrals between eigenstates of the $A$-particle and the
$(A-1)$-particle systems:
\begin{equation}
\phi _{\alpha }({\bf r})=\langle \Psi _{\alpha }^{(A-1)}|a({\bf
r})|\Psi ^{(A)}\rangle ,
\label{eq:OF}
\end{equation}
where $a({\bf r})$ is the annihilation operator for a nucleon
with spatial coordinate ${\bf r}$ (spin and isospin operators are
implied). In the mean-field approximation $\Psi ^{(A)}$ and $\Psi
_{\alpha }^{(A-1)}$ are single Slater determinants, and the
overlap functions are identical with the mean-field
single-particle wave functions, while in the presence of correlations both
$\Psi ^{(A)}$ and $\Psi _{\alpha }^{(A-1)}$ are complicated superpositions of
Slater determinants. In general, the overlap functions
$(\ref{eq:OF})$ are not orthogonal. Their norm defines the
spectroscopic factor
\begin{equation}
S_{\alpha }=\langle \phi _{\alpha }|\phi _{\alpha }\rangle .
\label{eq:spf}
\end{equation}
The normalized overlap function associated with the state $\alpha
$ then reads
\begin{equation}
\tilde{\phi}_{\alpha }({\bf r})=S_{\alpha }^{-1/2}\phi _{\alpha
}({\bf r}). \label{eq:NOF}
\end{equation}
The one-body density matrix (e.g. Eq. $(\ref{eq:obdm})$) can be
expressed in terms of the overlap functions in the form:
\begin{equation}
\rho ({\bf r},{\bf r^{\prime }})=\sum_{\alpha }\phi _{\alpha }^{*}({\bf r}%
)\phi _{\alpha }({\bf r^{\prime }})=\sum_{\alpha }S_{\alpha }\tilde{\phi}%
_{\alpha }^{*}({\bf r})\tilde{\phi}_{\alpha }({\bf r^{\prime }}).
\label{eq:rho}
\end{equation}

It is known (e.g. \cite{Mah91}) that the overlap function
associated with the bound state ($\alpha \equiv nlj$) of the
$(A-1)$- or $(A+1)$-nucleon system is an eigenstate of a
single-particle Schr\"{o}dinger equation in which the mass
operator plays the role of a potential. Due to its finite range,
the asymptotic behaviour of the radial part of the neutron
overlap functions for the bound states of the $(A-1)$-system is
given by \cite{Vn93,Bang85}:
\begin{equation}
\phi_{nlj}(r)\rightarrow C_{nlj}\exp(-k_{nlj}r)/r,
\label{eq:nasym}
\end{equation}
where $k_{nlj}$ is related to the neutron separation energy
\begin{equation}
k_{nlj}=\frac {\sqrt{2m\epsilon_{nlj}}}{\hbar}, \;\;\;
\epsilon_{nlj}=E_{nlj}^{(A-1)}-E_{0}^{A}.
\label{eq:decay}
\end{equation}
For proton bound states, due to an additional long-range part
originating from the Coulomb interaction, the asymptotic
behaviour of the radial part of the corresponding proton overlap
functions reads
\begin{equation}
\phi_{nlj}(r)\rightarrow C_{nlj}\exp[-k_{nlj} r-\eta \ln
(2k_{nlj}r)]/r,
\label{eq:pasym}
\end{equation}
where $\eta $ is the Coulomb (or Sommerfeld) parameter and $k_{nlj}$ in
(\ref{eq:decay}) contains in this case the mass of the proton and the proton
separation energy.

Taking into account Eqs. (\ref{eq:rho}) and (\ref{eq:nasym}), the
lowest ($n=n_{0}$) neutron bound-state $lj$-overlap function is
determined by the asymptotic behaviour of the associated partial
radial contribution of the one-body density matrix
$\rho_{lj}(r,r^{\prime})$ ($r^{\prime}=a\rightarrow \infty $) as
\begin{equation}
\phi _{n_{0}lj}(r)={\frac{{\rho _{lj}(r,a)}}{{C_{n_{0}lj}~\exp
(-k_{n_{0}lj}\,a})/a}}~,
\label{eq:nof}
\end{equation}
where the constants ${C_{n_{0}lj}}$ and ${k_{n_{0}lj}}$ are
completely determined by $\rho_{lj}(a,a)$. In this way the
separation energy $\epsilon_{n_{0}lj}$ and the spectroscopic
factor $S_{n_{0}lj}$ can be determined as well. Similar
expression for the lowest proton bound-state overlap function can
be obtained having in mind its proper asymptotic behaviour
(\ref{eq:pasym}).

The applicability of this theoretical procedure has been
demonstrated in Refs. \cite{Gai99,Gai2000,Sto96}. In particular,
it has been shown \cite{Gai2000} that the substantial realistic
inclusion of short-range as well as tensor correlations in the
ODM of $^{16}$O constructed within the Green Function Method
\cite{Po96} and, as a consequence, in the extracted overlap
functions, leads to a fair and consistent description of the
experimental cross sections of the $^{16}$O$(e,e^{\prime}p)$ and
$^{16}$O$(\gamma,p)$ reactions. The importance of collective long-range
correlations on the additional spectroscopic factors extracted in
comparison with data has been also pointed out in \cite{Gai2000}.

In this work we use as a basis the GCM results for the correlated
states of $^{40}$Ca obtained in \cite{Iva2000}, where
correlations due to collective nucleon motion are properly taken
into account. The procedure to calculate the overlap functions
from the one-body density matrix outlined above is performed
numerically for $^{40}$Ca within the GCM. In this respect we
would like to note that an important condition of this procedure
is the exponential asymptotics of the overlap functions at
$r^{\prime}\rightarrow \infty$ (see Eqs. (\ref{eq:nasym}) and
(\ref{eq:pasym})), which is related to the correct asymptotics of
$\rho_{lj}(r,r^{\prime})$ at $r^{\prime}\rightarrow \infty$. This
condition is fulfilled in our GCM approach by using realistic
single-particle Woods-Saxon wave functions in the generating
single Slater determinant wave function
${\Phi}(\{{\bf{r}}_i\},x)$ in Eq. (\ref{eq:psi}). The exponential
decay of the partial radial ODM is with a decay constant
$k_{nlj}$ related to the separation energy (Eq.
(\ref{eq:decay})). The question about the correct asymptotic
behaviour of the extracted overlap functions will be discussed
also in Section III.

\section{Results and discussion}
In the present paper the GCM numerical calculations have been
performed using Woods-Saxon construction potential with
diffuseness parameter as a generator coordinate. The values of
the radius and the depth of the construction potential have been
fixed. Following \cite{BH91}, the value of the depth used for
both neutron and proton cases has been taken to be 50 MeV, while
the corresponding values of the potential radius are 1.30 fm for
the neutron and 1.24 fm for the proton case, respectively. The
effective $SkM^{*}$ interaction \cite{Bar82} is used, with
parameter set values ($t_{0}=-2645, t_{1}=410, t_{2}=-135, t_{3}=
15595, {\sigma}=1/6$), which give the realistic binding energy of
$^{40}$Ca obtained from the Hill-Wheeler equation (\ref{eq:HW}).
A discretization procedure is performed in order to solve this
integral equation using a set of regular mesh points with steps
as well as ranges of values of the generator coordinate which
lead to the convergence of the results, i.e. the latter do not
change after decreasing the step size or increasing the range of
the generator coordinate value. The range of variation of the
diffuseness parameter turns out to be within the interval
0.49$\div $0.9072 fm. The ground-state energy and the excitation
energy of the first monopole $0^{+}$ state of $^{40}$Ca derived
from the procedure (340.07 MeV and 20.4 MeV) are close to the
corresponding experimental values (342.06 MeV and 20.0 MeV
\cite{Voi85}, respectively). Likewise, in \cite{Iva2000} the same
GCM scheme gives a value for the excitation energy of the $0^{+}$
breathing state in $^{16}$O which is in a good agreement with the
result obtained in \cite{Bri63}.

The one-body density matrix (\ref{eq:obdm}) constructed within the
GCM approach has been applied to calculate overlap functions
related to the quasihole states in the $^{40}$Ca nucleus. In our
method the extracted overlap functions for a given orbital
momentum $l$ are the same for the different $j=l\pm 1/2$.
Hereafter the overlap functions for the neutron and proton $1d$
and $2s$ states will be of particular interest: transitions to
the $3/2^{+}$ ground state and the first $1/2^{+}$ excited state
of the corresponding residual nuclei observed in
$^{40}$Ca($e,e'p$), $^{40}$Ca($\gamma,p$) and $^{40}$Ca($p,d$)
reactions will be considered. The squared overlap functions
$|r\phi(r)|^2$ for the neutron and proton $1d$ and $2s$ quasihole
states are illustrated in Fig. \ref{fig:ovp}. They have the
correct asymptotic behavior related to the exponential decrease
of the ODM in its asymptotic region, being different for protons
and neutrons. The latter is in accordance with the experimental
proton and neutron separation energies. As known, the one-neutron
separation energy for $^{40}$Ca is almost twice larger than the
one-proton one and, therefore, the asymptotic tail of the GCM
density matrix is dominated by the contribution from the proton
overlap between the $^{40}$Ca and $^{39}$K ground-state wave
functions, where the role of the Coulomb interaction is important.

The comparison of $|r\phi(r)|^2$ calculated with the GCM overlap
functions and with the shell-model Woods-Saxon wave functions for
the neutron bound states is given in Fig. \ref{fig:comp}. Small
differences between the two functions can be seen for both $1d$
and $2s$ states. The same is valid also for the $1s$ and $1p$
quasihole states, not presented in the figure. This result
justifies the approximation of using shell-model orbitals instead
of overlap functions in calculating the nucleon knockout cross
section for these nuclear states. Nevertheless, as it has been
pointed out in previous works \cite{Gai99,Gai2000}, the
calculated cross sections of one-nucleon removal reactions are
very sensitite to the differences in the corresponding overlap
functions.

The values of the neutron and proton separation energies derived
from the procedure mentioned above are listed in Table 1. It can
be seen that the separation energies obtained within the GCM are
very close to the experimental values for the $1d$ state of
$^{40}$Ca and are anyhow better than the JCM result  for
$\epsilon_{n}$ in the $2s$ state. In addition, we present the
separation energy values obtained within the mean-field
approximation (the Hartree-Fock method \cite{MS91}). We should
note that the differences between the latter and those from both
correlation methods are dependent on whether the nucleons are in
a valence orbital or not. Here we would like to note also that the
use of Woods-Saxon wave functions (instead of e.g.
harmonic-oscillator wave functions as in the JCM calculation),
which have the proper exponential-type decay, leads to a correct
asymptotic behaviour of the extracted overlap functions with
satisfactory separation energies of the quasihole states.

The values of the spectroscopic factors deduced from the
numerical procedure turn out to be very close to unity. This is
not surprising, because tensor and short-range nucleon-nucleon
correlations are not included in the GCM. As known, these correlations
are responsible for the bulk part of the depletion of the occupied
states \cite{Gai99}. The values of the spectroscopic
factors derived from the Jastrow correlation method \cite{Sto96}, where
short-range correlations are included, are 0.892 for the $1d$ state and
0.956 for the $2s$ state, respectively.

The reduced cross sections for the $^{40}$Ca($e,e'p$) reaction as
a function of the missing momentum $p_{m}$, i.e. the magnitude of
the recoil momentum of the residual nucleus, for the transitions
to the $3/2^{+}$ ground state and the first $1/2^{+}$ excited
state [at excitation energy $E_{x}$=2.522 MeV] of $^{39}$K are
displayed in Figs. \ref{fig:eepgr} and \ref{fig:eepex},
respectively. Calculations have been done with the code DWEEPY
\cite{DWEEPY}, which is based on a nonrelativistic distorted wave
impulse approximation (DWIA) description of the nucleon knockout
process and includes final-state interactions and Coulomb
distortion of the electron waves \cite{Oxford}. In the figures
the results obtained with the overlap functions generated by the
GCM are compared with the data from \cite{Kra90} in parallel and
perpendicular kinematics. In parallel kinematics the momentum of
the outgoing proton is fixed and is taken parallel or
antiparallel to the momentum transfer. Different values of the
missing momentum are then obtained by varying the electron
scattering angle. In perpendicular kinematics the outgoing proton
energy and the momentum transfer are kept constant and different
values of the missing momentum are obtained by varying the angle
of the outgoing proton.

A fair agreement with the shape of the experimental cross section
is achieved for both transitions and kinematics with the overlap
functions emerging from the ODM calculated within the GCM, which
includes realistic correlations corresponding to single-particle
and collective motion modes. The OF's already include the
spectroscopic factors. In order to reproduce the size of the
experimental cross section, an additional reduction factor has
been applied to the theoretical results. As in our previous
analysis of the $^{16}$O($e,e'p$) reaction \cite{Gai2000}, this
factor can be considered as a further spectroscopic factor,
reflecting the correlations not included in the OF which
correspondingly cause depletion of the quasihole states. The same
reduction factor has been applied to the cross sections
calculated with the OF for both parallel and perpendicular
kinematics, namely 0.55 for the $3/2^{+}$ ground state in Fig.
\ref{fig:eepgr} and 0.50 for the $1/2^{+}$ excited state in Fig.
\ref{fig:eepex}. Small variations around these values do not
change significantly the comparison with the data. However, for
the ground-state transition in perpendicular kinematics a
reduction factor about 20\% lower would give a better agreement
with data. A different spectroscopic factor for this transition
in the two kinematics is found also in the data analysis of
\cite{Kra90}, where in perpendicular kinematics the spectroscopic
factor is 25\% lower than the one determined in parallel
kinematics. This discrepancy is related in \cite{Kra90} to the
basic ingredients of the theory: the optical potential, the
bound-state wave function, the electron distortion and the
influence of meson-exchange currents (MEC). Our total
spectroscopic factors (the norm of OF's multiplied by the
additional reduction factor mentioned above) with the GCM, 0.54
for $3/2^{+}$ and 0.49 for $1/2^{+}$, are comparable with the
"experimental" ones, 0.65 and 0.52, determined under parallel
kinematical conditions in the analysis of \cite{Kra90}, where the
calculations are performed within the same DWIA framework and
with the same optical potential, but where phenomenological
single-particle wave functions are used with some parameters
adjusted to the data. We emphasize that in the present analysis
the overlap functions theoretically calculated on the basis of the
ODM of $^{40}$Ca within the GCM do not contain free parameters.
Our spectroscopic factors are also in a reasonable agreement with
those extracted from the fully relativistic theoretical analysis
of the $^{40}$Ca($e,e'p$) reaction in \cite{Udi93}, where,
however, only a part of the ($e,e'p$) data from \cite{Kra90} is
considered, while somewhat larger spectroscopic factors (around
0.7--0.8) are obtained in the relativistic analyses of
\cite{Jin92,Jo96}.

In Fig. \ref{fig:eepgr} for the
$^{40}$Ca($e,e'p$)$^{39}$K$_{g.s.}$ reaction we present also the
results obtained with the phenomenological Woods-Saxon wave
function. The WS wave function is also able to give a good
description of the data with a reduction factor of 0.6625,
somewhat larger than the one applied to the reduced cross section
computed with GCM overlap function. Also with the WS wave
function a reduction factor lower by about 20\% would give a
better description of the data in perpendicular kinematics.

Fig. \ref{fig:gp60} shows the angular distribution of the
$^{40}$Ca($\gamma,p$)$^{39}$K$_{g.s.}$ reaction at
$E_{\gamma}$=60 MeV. In the figure the results given by the sum
of the one-body and of the two-body seagull currents are compared
with the contribution given by the one-body current, which roughly
corresponds to the DWIA treatment based on the direct knockout
(DKO) mechanism. Both contributions of DKO and MEC are
consistently evaluated in the theoretical framework of
ref.\cite{Benenti}. The results obtained with the OF from GCM for
the ground state transition and with the phenomenological WS wave
function are compared in the figure. In order to check the
consistency in the description of different one-proton removal
reactions, the calculated cross sections have been multiplied by
the same reduction factors obtained from the analysis of
($e,e'p$) data, i.e. 0.55 with GCM and 0.6625 with WS. The
differences between the two curves are considerable and larger
than in the ($e,e'p$) reaction. This result might have been
expected, since it is well known that ($\gamma,p$) results are
extremely sensitive to the theoretical ingredients adopted in the
calculation, in particular for the bound states.

The GCM calculation lies well below the data in DWIA, but a
reasonable agreement with the size and the shape of the
experimental cross section is obtained when MEC are added. Thus,
the important role played by MEC \cite{Gai2000,Benenti} is
confirmed here also for the $^{40}$Ca($\gamma,p$) reaction. The
WS calculation are larger that the ones with the GCM. This result
can be understood also from Fig. \ref{fig:eepgr}, where at the
high values of the missing momentum, that are explored in the
($\gamma,p$) reaction, the ($e,e'p$) reduced cross sections
calculated with WS wave function are larger than the ones
obtained with the OF from the GCM. Thus, in DWIA the WS result is
closer to the data, in particular for the lowest angles, but when
MEC are added it overshoots the experimental cross section.

Therefore, although both calculations with the GCM and WS wave
functions are able to give a good description of the
$^{40}$Ca($e,e'p$) data for the transition to the $3/2^{+}$
ground state of $^{39}$K, the ($\gamma,p$) results presented in
Fig. \ref{fig:gp60} for the same transition show that the GCM
overlap function leads to a better and more consistent description
of data for the $(e,e'p)$ and $(\gamma,p)$ reactions. This result
suggests proper accounting for the nucleon correlation effects in
the framework of the GCM.

\begin{figure}
\caption[]{Differential cross section for the $^{40}$Ca($p,d$)
reaction at incident proton energy $E_{p}=65$ MeV for the
transition to the $3/2^{+}$ ground state in $^{39}$Ca. The neutron
overlap function is derived from the ODM of GCM (solid line). The
proton and deuteron optical potentials are taken from refs.
\cite{Ful69} and \cite{Becc69}, respectively. The experimental
data \cite{Roos75} are given by the full circles. The theoretical
results have been multiplied by the reduction factor 0.55.
\label{fig:pd} }
\end{figure}
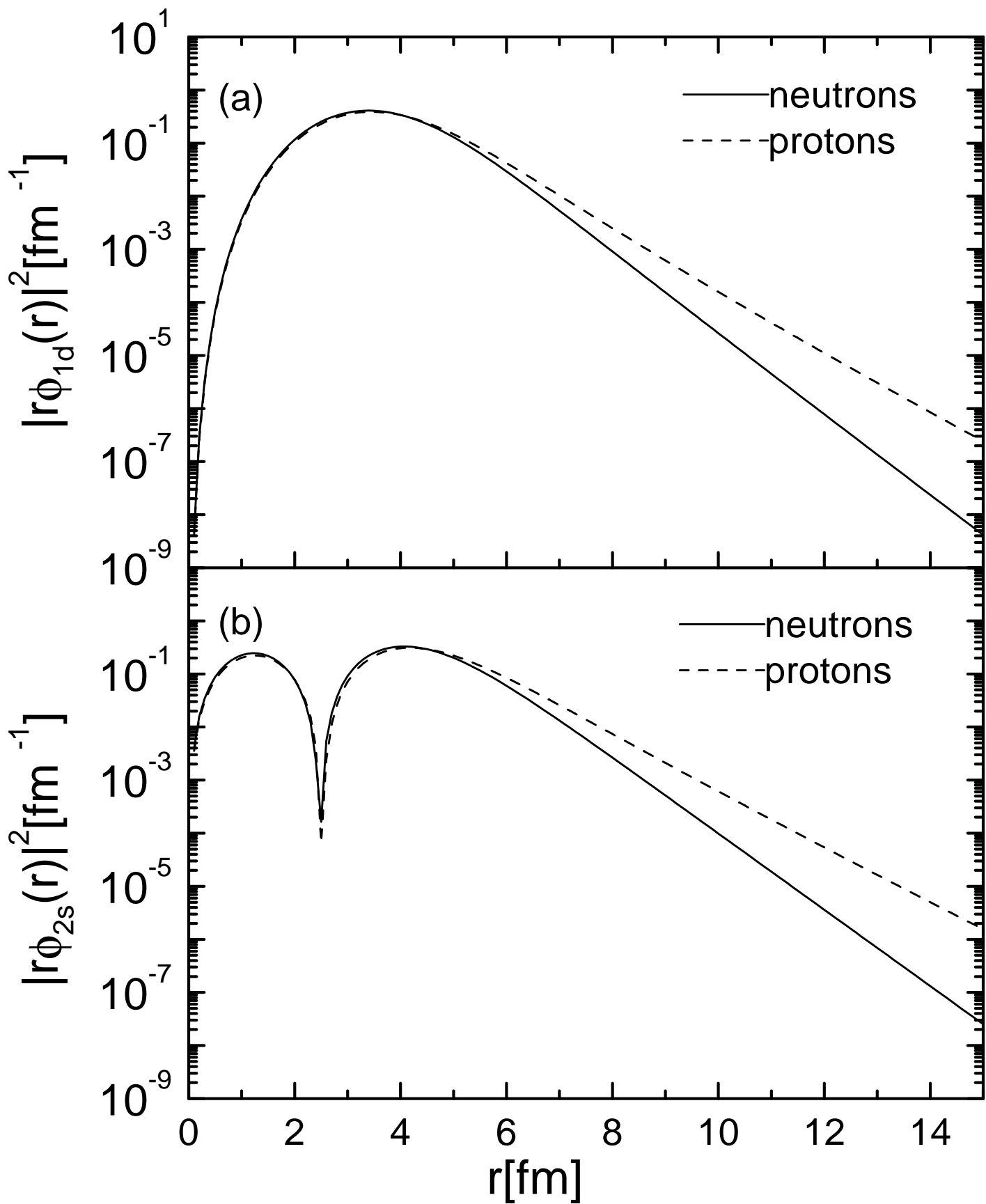
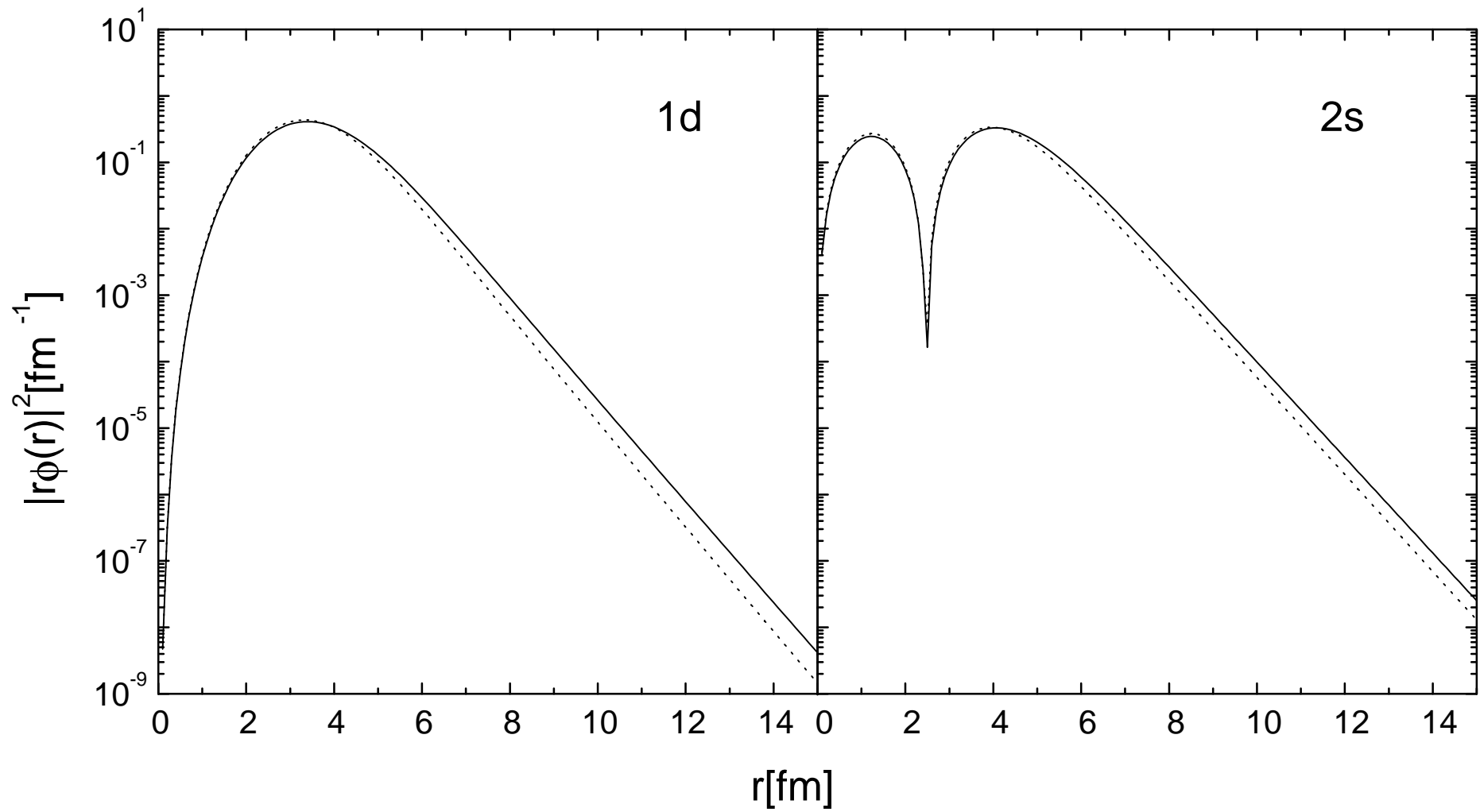
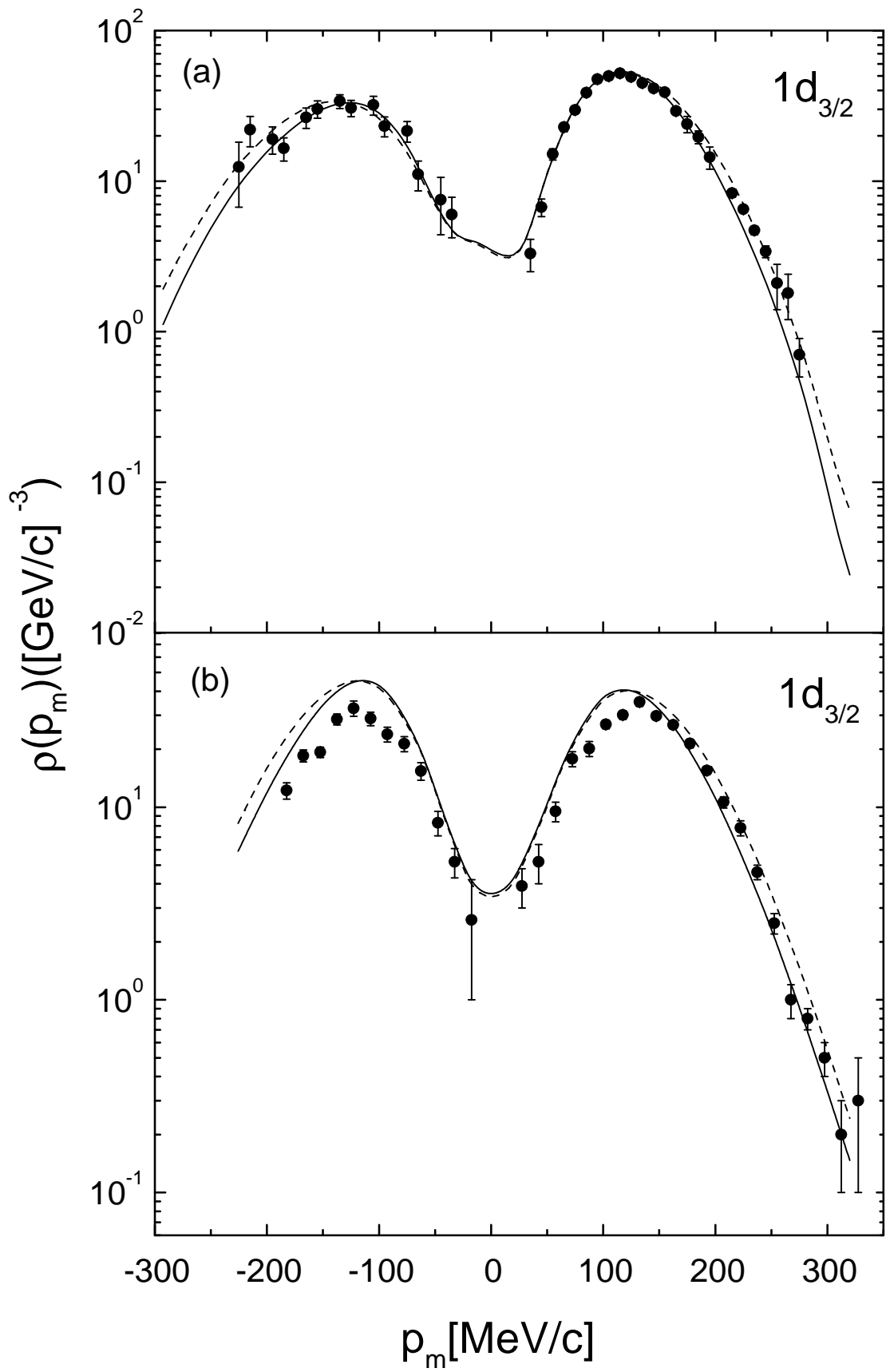
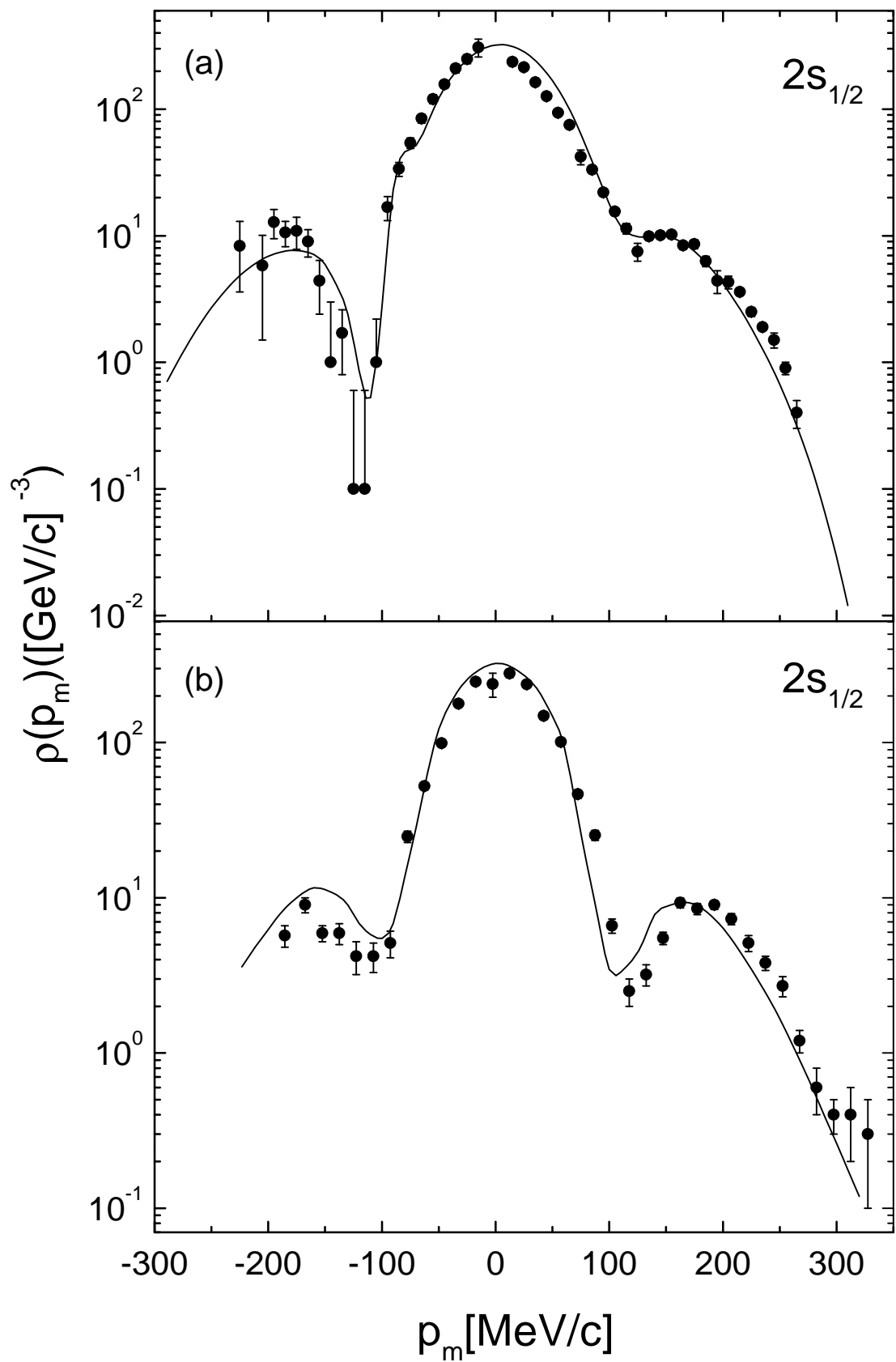
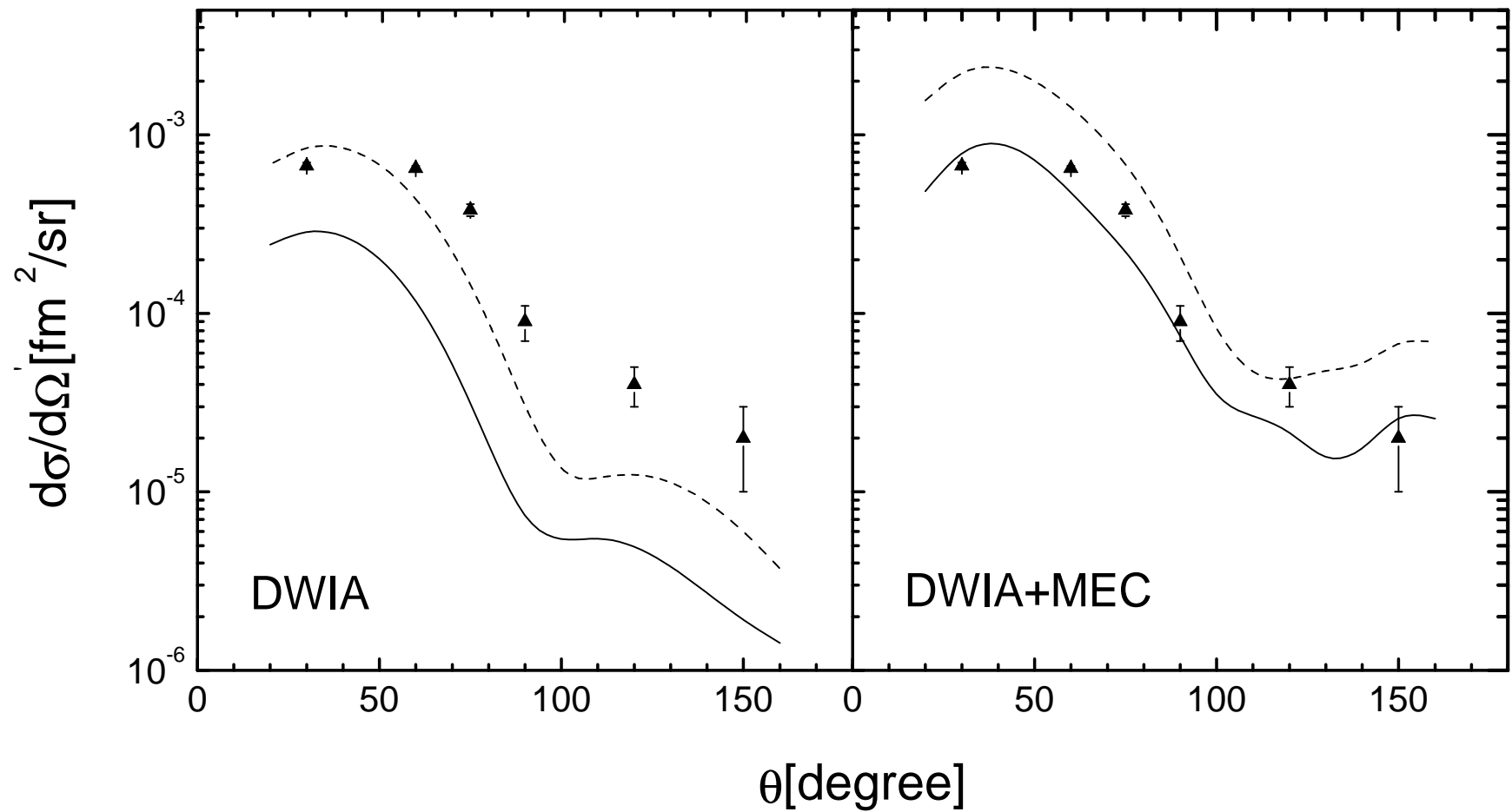
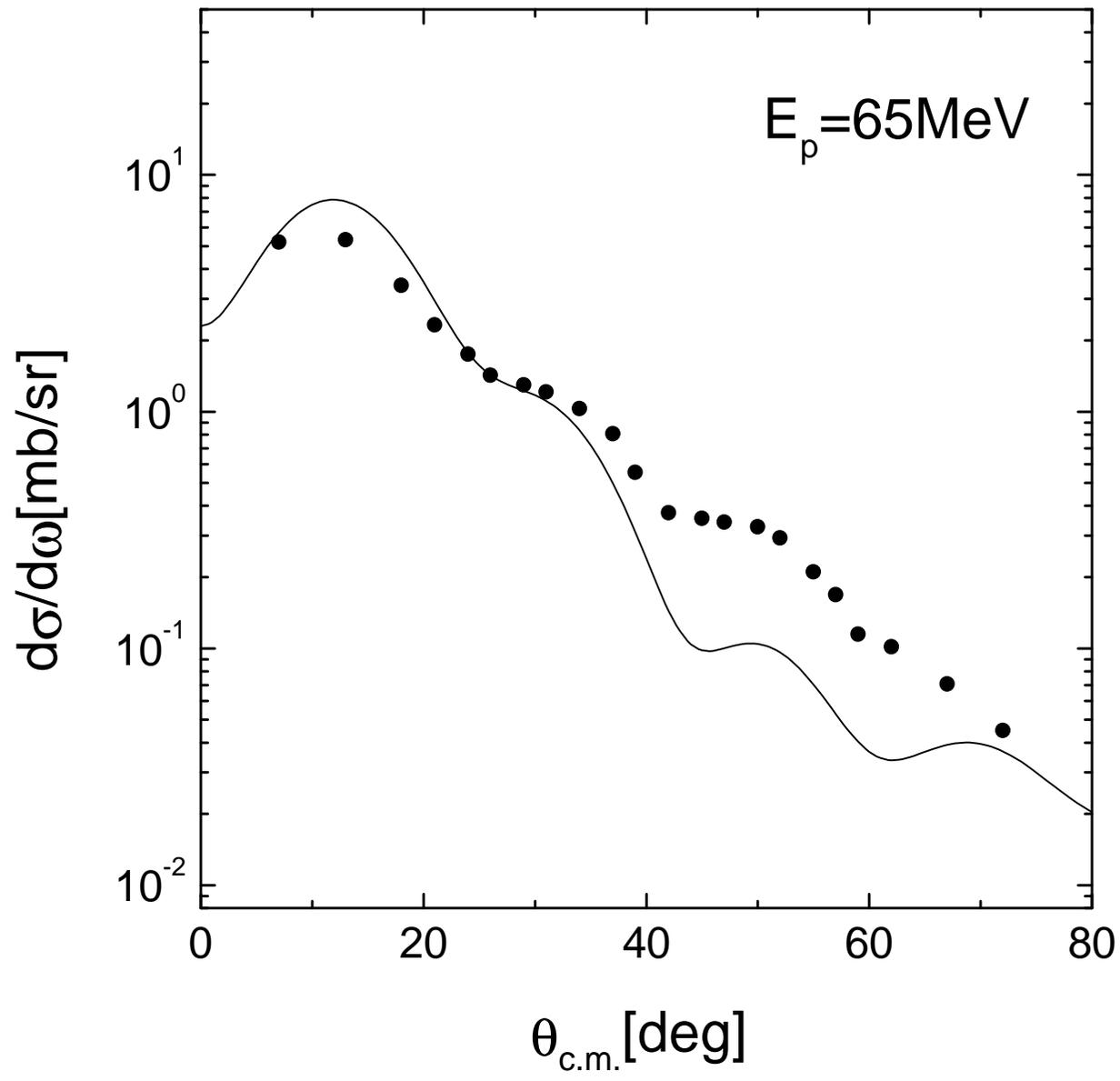

In Fig. \ref{fig:pd} the differential cross section of the
$^{40}$Ca($p,d$) reaction for the transition to the $3/2^{+}$
ground state in $^{39}$Ca, calculated at incident proton energy
$E_{p}$=65 MeV with the neutron GCM overlap function, is given and
compared with the experimental data \cite{Roos75}. It has been
obtained using the Distorted Wave Born Approximation (DWBA) with
zero-range approximation for the $p$-$n$ interaction inside the
deuteron. It is well known that the cross section of the ($p,d$)
reaction is more sensitive to the reaction mechanism adopted than
to the choice of the bound-state wave function. In this respect
quasifree nucleon knockout reactions are more suitable for our
investigation. Nevertheless, within the DWBA for such a pick-up
process as the ($p,d$) reaction, it can be seen that the shape of
the angular distribution is well reproduced by the GCM. Applying
the total spectroscopic factor of 0.54 for $3/2^{+}$ state to the
calculated cross section, a reasonable agreement with the size of
the experimental cross section is achieved. The role of the
additional depletion or reduction of the spectroscopic factors
due to correlations not included in the OF has been already seen
on the examples of the previous reactions considered. Although
some differences between the calculated and experimental angular
distributions exist, we would like to emphasize that, in
principle, overlap functions extracted from realistic one-body
density matrices have to be used as a tool for an accurate
description of one-nucleon removal reactions.

\section{Conclusions}
In summary, we have calculated single-particle overlap functions
and the spectroscopic factors corresponding to low-lying
(quasihole) $(A-1)$-nucleon states on the base of the one-body
density matrix of $^{40}$Ca by considering its asymptotic (large
$r$) region and using the Generator Coordinate Method. These
calculations have been performed within an approach in which a
Woods-Saxon construction potential and a Skyrme-type effective
force are used. A consistent analysis of the cross sections of
one-nucleon removal reactions on $^{40}$Ca by means of the same
overlap functions is given. In contrast to standard DWIA
analyses, where phenomenological single-particle wave functions
were used with some parameters fitted to the data, in this paper
the results have been obtained with theoretically calculated
overlap functions which do not include free parameters. The
differences between the GCM overlap functions and the shell model
WS wave functions for both hole- and particle- states are analyzed
as a test for the role of correlations inherent to our approach.

We have found that the calculated cross sections are generally in
good agreement with the experimental data. The quality of the
agreement, however, is sensitive to details of the overlap
functions. The important role of the additional reduction
spectroscopic factor applied to the present calculations is
pointed out. The fact that the reduction factor is consistent in
different nucleon removal reactions and also that this result is
obtained both for $^{40}$Ca in the present work and for $^{16}$O
in \cite{Gai2000} gives a more profound theoretical meaning to
this parameter. It can be interpreted as the spectroscopic factor
accounting for correlations not included in the Generator
Coordinate Method. The values of the total spectroscopic factors
obtained in the present analysis are in reasonable agreement with
those given by previous theoretical investigations.

Apart from the short-range and tensor correlations studied in
previous papers \cite{Gai99,Gai2000}, in this work we looked into
the role of correlations caused by the collective nucleon motion.
Exploring the GCM, a consistent picture for all three $(e,e'p)$,
$(\gamma,p)$ and $(p,d)$ reactions on $^{40}$Ca was achieved on
the same theoretical ground. The results indicate that the
effects of NN correlations taken into account within our approach
and which are of long-range type are of significant importance
for the correct analysis of the processes considered.

Finally, we would like to emphasize that the theoretical method to
calculate the cross sections of one-nucleon removal reactions by
means of single-particle overlap functions for the bound states
presented in this paper is, in principle, valid for all kind of
nuclei. The particular GCM scheme employed in our work is
applicable, however, only to nuclei with equal numbers of protons
and neutrons but to both closed and non-closed shell nuclei.
Calculations of similar reactions on open $s$-$d$ shell nuclei
are in progress.

\acknowledgments We acknowledge the financial support given by
the Bulgarian National Science Foundation under Contracts
$\Phi$--809 and $\Phi$--905.

\newpage

\noindent {\bf Table 1:} Neutron ($\epsilon_{n}$) and proton
($\epsilon_{p}$) separation energies (in MeV) calculated on the
basis of the one-body density matrix for $^{40}$Ca within the
GCM. Comparison is made with the Hartree-Fock (HF) \cite{MS91}
single-particle energies, JCM results \cite{Sto96} and with the
experimental data (EXP) \cite{MS91}. The HF and EXP values are
given for the $d_{3/2}$ state.
\vspace{1cm}

\begin{center}
\begin{tabular}{cccccccccc}
\hline\hline
& & \multicolumn{4}{c}{$\epsilon_{n}$} & &  \multicolumn{3}{c}{$\epsilon_{p}$} \\
\cline{3-6} \cline{8-10}
nl & & GCM   & HF  & JCM & EXP & & GCM   & HF  & EXP \\
\hline
1d  & & 15.84 & 17.52 & 24.75 & 15.64 & &  8.69 & 10.33 & 8.33 \\
2s  & & 14.15 & 19.51 & 13.07 & 18.19 & &  7.39 & 12.42 & 10.94\\
\hline\hline
\end{tabular}
\end{center}

\newpage
\begin{figure}
\caption[]{$|r\phi(r)|^2$ for the neutron and proton $1d$ (a) and
$2s$ (b) quasihole states in $^{40}$Ca obtained with GCM. The
normalization is: $\int \phi^{2}(r)r^{2}dr=S$. \label{fig:ovp} }
\end{figure}

\begin{figure}
\caption[]{$|r\phi(r)|^2$ for the neutron $1d$ and $2s$ quasihole
states in $^{40}$Ca obtained with GCM overlap function (solid
line) and Woods-Saxon single-particle wave function
$|r\varphi(r)|^2$ (dashed line). All curves are normalized to
unity. \label{fig:comp} }
\end{figure}

\begin{figure}
\caption[]{Reduced cross section of the $^{40}$Ca($e,e'p$)
reaction as a function of the missing momentum $p_{\mathrm m}$
for the transition to the $3/2^{+}$ ground state of $^{39}$K in
parallel (a) and perpendicular (b) kinematics. The incident
electron energy is 460 MeV in case (a), 483.2 MeV in case (b) and
the outgoing proton energy is 100 MeV. The optical potential is
taken from ref. \cite{Schwandt}. The overlap function is derived
from the ODM of GCM (solid line). The dashed line is calculated
with the WS wave function. The experimental data (full circles)
are taken from ref.\cite{Kra90}. The theoretical results have been
multiplied by the reduction factors given in the text.
\label{fig:eepgr} }
\end{figure}

\begin{figure}
\caption[]{The same as in Fig. 3 for the transition to the first
$1/2^{+}$ excited state at 2.522 MeV of $^{39}$K. The theoretical
results have been multiplied by the reduction factor 0.50.
\label{fig:eepex} }
\end{figure}

\begin{figure}
\caption[]{Angular distribution of the $^{40}$Ca($\gamma,p$)
reaction for the transition to the $3/2^{+}$ ground state of
$^{39}$K at $E_\gamma = 60$ MeV. The optical potential is taken
from ref. \cite{Schwandt}. The separate contributions given by
the one-body current (DWIA) and the final result given by the sum
of the one-body and the two-body seagull current (DWIA+MEC) are
shown. Line convention is as in Fig. \ref{fig:eepgr}. The
experimental data (triangles) are taken from ref.\cite{Abe92}.
The theoretical results have been multiplied by the reduction
factors given in the text, consistent with the analysis of
($e,e'p$) data. \label{fig:gp60} }
\end{figure}

\end{document}